\documentclass[sigplan,screen]{acmart}
\setcopyright{acmlicensed}
\acmPrice{15.00}
\acmDOI{10.1145/3425898.3426952}
\acmYear{2020}
\copyrightyear{2020}
\acmSubmissionID{gpce20main-p2-p}
\acmISBN{978-1-4503-8174-1/20/11}
\acmConference[GPCE '20]{Proceedings of the 19th ACM SIGPLAN International Conference on Generative Programming: Concepts and Experiences}{November 16--17, 2020}{Virtual, USA}
\acmBooktitle{Proceedings of the 19th ACM SIGPLAN International Conference on Generative Programming: Concepts and Experiences (GPCE '20), November 16--17, 2020, Virtual, USA}

\usepackage{booktabs}   %% For formal tables:
\usepackage{subcaption} %% For complex figures with subfigures/subcaptions
\usepackage{microtype}
\usepackage{multirow}
\usepackage[frozencache,cachedir=.]{minted}
\usepackage{array}
\usepackage{commath}
\usepackage{bm}
\usepackage{tabu}
\usepackage{longtable}

\usepackage{tabularx}
\newcolumntype{Y}{>{\centering\arraybackslash}X}

\usepackage{enumitem}
\usepackage{listings}
\usepackage{subcaption}

\usepackage{cleveref}

\usepackage{tikz}
\usepgflibrary{fpu}

\usepackage{siunitx}

\usepackage{pifont}% http://ctan.org/pkg/pifont

\newcommand{\makespeare}{\textsc{Makespeare}}
\newcommand{\simpl}{\textsc{Simpl}}
\newcommand{\LL}{$\bm{\lambda^2}$}
\newcommand{\sketchadapt}{\textsc{SketchAdapt}}
\newcommand{\baseline}{\textsc{TypeDirect}}
\newcommand{\presyn}{\textsc{Presyn}}
\newcommand{\IID}{IID}
\newcommand{\markov}{Markov}

\newcommand{\bfsc}[1]{\textbf{\textsc{#1}}}

\mathchardef\mhyphen="2D

\usepackage{balance}

\begin{document}

\title{Modeling Black-Box Components with Probabilistic Synthesis}
%Probabilistic Black-Box Program Synthesis}

\author{Bruce Collie}
\orcid{0000-0003-0589-9652}
\affiliation{
  \department{School of Informatics}
  \institution{University of Edinburgh}
  \streetaddress{10 Crichton Street}
  \city{Edinburgh}
  \postcode{EH8 9AB}
  \country{United Kingdom}
}
\email{bruce.collie@ed.ac.uk}

\author{Jackson Woodruff}
\affiliation{
  \department{School of Informatics}
  \institution{University of Edinburgh}
  \streetaddress{10 Crichton Street}
  \city{Edinburgh}
  \postcode{EH8 9AB}
  \country{United Kingdom}
}
\email{J.C.Woodruff@sms.ed.ac.uk}

\author{Michael F.P.\ O'Boyle}
\orcid{0000-0003-1619-5052}
\affiliation{
  \department{School of Informatics}
  \institution{University of Edinburgh}
  \streetaddress{10 Crichton Street}
  \city{Edinburgh}
  \postcode{EH8 9AB}
  \country{United Kingdom}
}
\email{mob@inf.ed.ac.uk}

%% Abstract
%% Note: \begin{abstract}...\end{abstract} environment must come
%% before \maketitle command
\begin{abstract}
This paper is concerned  with synthesizing programs based on \emph{black-box}
oracles: we are interested in the case where there exists an executable
implementation of a component or library, but its internal structure is unknown.
We are provided with just an API or function signature, and aim to synthesize a
program with equivalent behavior.

To attack this problem, we detail \presyn{}: a program synthesizer designed for
flexible interoperation with existing programs and compiler toolchains.
\presyn{} uses high-level imperative control-flow structures and a pair of
cooperating predictive models to efficiently narrow the space of potential
programs. These models can be trained effectively on small corpora of
synthesized examples.

We evaluate \presyn{} against five leading program synthesizers on a collection
of 112 synthesis benchmarks collated from previous studies and real-world
software libraries. We show that \presyn{} is able to synthesize a wider range
of programs than each of them with less human input. We demonstrate the
application of our approach to real-world code and software engineering problems
with two case studies: accelerator library porting and detection of duplicated
library reimplementations.

\end{abstract}



%% Keywords
%% comma separated list
  %% \keywords are mandatory in final camera-ready submission

%% \maketitle
%% Note: \maketitle command must come after title commands, author
%% commands, abstract environment, Computing Classification System
%% environment and commands, and keywords command.
\maketitle

\begin{figure*}
  \centering
  \includegraphics[width=\textwidth]{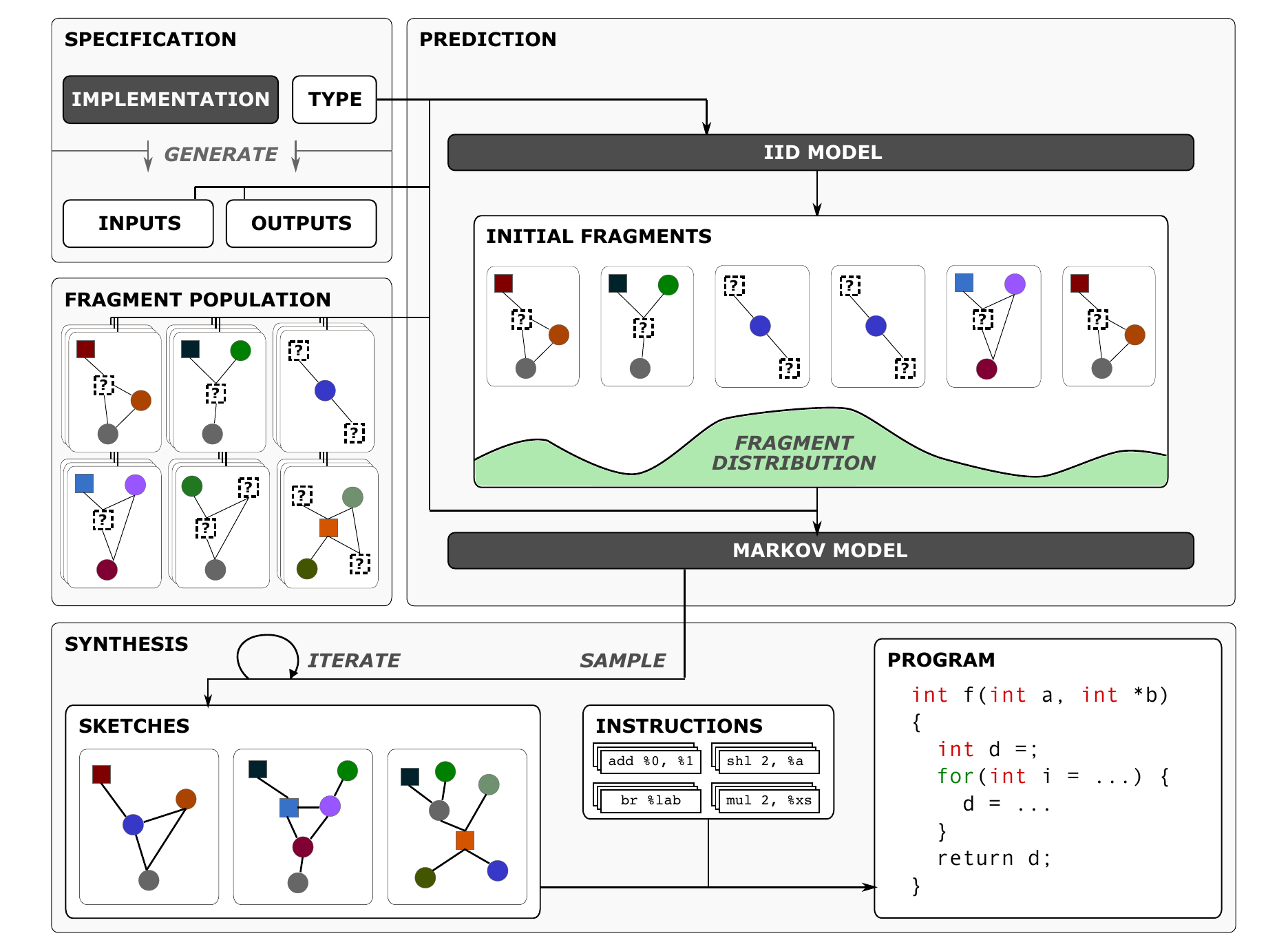}
  \caption{
    A summary of \presyn{}'s implementation. Fragment distributions are learned
    by \IID{} and \markov{} models using the problem specification and initial
    fragment population. Sketches are then sampled and synthesized into
    executable programs.
  }
  \label{fig:system}
\end{figure*}

\section{Introduction} \label{sec:intro}
Modeling and understanding the behavior of software components is a key issue in
software engineering \cite{Cambronero2019}. It has been used to discover
similarity between libraries \cite{Teyton2013a}, rejuvenate code
\cite{Mendis2015} and  match accelerators to software \cite{Collie2019}.  While
a formal model of a  component may be available, it is often supplied as
low-level binary  or not at all \cite{Mendis2015}.
% existing work in software engineering seeks to minimize the reliance on such
% models \cite{Phan2017} when performing transformations.
Our goal is to model such "black-box" software components, in a form suitable
for interoperation with real world code using \emph{program synthesis}.

For the sake of wide applicability, we make minimal assumptions about the
component being modeled: we require only that it exists in an executable form
and has a known type signature. This paper therefore addresses the problem of
program synthesis based on the observed behavior of an oracle performing an
existing, but unknown computation \cite{jha2017theory}. We have no knowledge of
the oracle's internal structure, but capturing its behavior in the form of
input-output (IO) examples is inexpensive.

This problem falls under the domain of \emph{programming by example} which has
received considerable interest from industry \cite{gulwani2016programming}.
Here, the aim is to synthesize a program from user provided examples
\cite{Gulwani2011,An2019}. Our approach differs from the standard formulation in
that IO examples are effectively free, as we do not rely on the user.
% For this reason we are able
% to tackle more complex  tasks than, for example,  data wrangling
% \cite{gulwani2016programming}.
 
Our approach is distinct to much prior work, where their aim is to generate
provably correct programs with respect to a formal specification (typically
making use of counter-examples and SMT solvers)
\cite{abate2018counterexample,Butler2017,Chu2017}. Instead, it is closer in
spirit to neural synthesis approaches \cite{Parisotto2016a,Balog2016,Nye2019},
where IO examples {\em are} the specification of a task. These schemes, however,
are currently limited in the problem domains that can be addressed.
\cite{polgreen2020counterexample}. 

% Conversely, the well established syntax guide program syntheis competition,
% SyGuS \cite{SyGuS}  is limited to strings, bitvectors or linear integer
% arithmetic where  the functions synthesized have a single return value and the
% functions are loop-free.

Like previous work which attempts to synthesize without a formal specification,
we cannot guarantee correctness of the generated program \cite{An2019}.
However, testing over a large set of examples can provide observational
equality, which is widely accepted as a sufficiently strong guarantee
\cite{Claessen2000} in the absence of formal proof.

% In our experiments we find all synthesizers produce correct results and  in practice, 
% developers could be asked to sign off on  synthesized programs.

\paragraph{Sketching}

Our approach uses ideas from \emph{sketching} program synthesis
\cite{Solar-Lezama2009}: a high level partial structure, or sketch, of the
target program allows for efficient search through a space of potential
solutions. While many techniques use externally provided sketches \cite{So2017},
our scheme uses a two-phase synthesis process \cite{Wang2017} where sketches are
constructed by the synthesizer based on the problem specification (in our case,
a type signature and IO examples).

However, IO examples alone are not sufficient to synthesize programs that use
complex components. We use prior observations of synthesized program structure
to build a probability distribution over fragments of sketches given a type
signature and a set of IO examples. Specifically, we develop two models: \IID{}
and \markov{}.

The first, \IID{}, assumes independence among sketch fragments ({\bf
I}ndependent and {\bf I}dentically {\bf D}istributed), while the second assumes
the next fragment is dependent on the current one (i.e.\ a Markov model).  Using
this approach, we synthesize a wide range of complex components from
black-box specifications.

There is a large body of related work in this diverse area.  and providing a
fair comparison is a challenge \cite{Helmuth2015,Pantridge2017}. In this paper
we attempt a fair, systematic and reproducible evaluation of representative
state-of the art existing schemes. 

As our approach is driven by a probabilistic model of sketches, we analyze the
learned distributions and the resulting insights into program structure. We then
evaluate the use of black-box synthesis in two application areas: rejuvenating
legacy scientific code and detecting reimplementations of library code.

\subsection{Contributions}

We implement a novel program synthesizer (\presyn{}) that uses lightweight
probabilistic models to efficiently search for solutions in varied problem
domains. We evaluate \presyn{} against five other program synthesizers from
different traditions: neural \sketchadapt{} \cite{Nye2019}, functional \LL{}
\cite{Feser2015}, imperative \simpl{} \cite{So2017}, type-directed \baseline{}
\cite{Collie2019} and genetic \makespeare{} \cite{Rosin2018},

This paper makes the following main contributions:
\begin{itemize}
\item Probabilistic synthesis based on corpus priors with broader synthesis
  results than existing work
\item An extensive and systematic evaluation of example-based synthesis from
  different research domains
\item An analysis of sketch distributions across synthesized programs
\item Two case studies showing uses of black box synthesis for software
  engineering
\end{itemize}

\section{Overview} \label{sec:overview}
\Cref{fig:system} gives an overview of our program synthesizer, \presyn{}. It
consists of four primary components: specification and example construction;
fragment population; probabilistic fragment models and program synthesis.

Given a function signature, we can generate inputs by sampling random values
from the domain of each input parameter. As we have an executable implementation
of the component, we evaluate it on these inputs and record its output behavior
(i.e.\ the function return value and any writes to memory). The function
signature and IO examples form the component's specification
(\bfsc{specification} in \Cref{fig:system}).

The next step is to predict which fragments from a population (\bfsc{fragment
population}) are most likely to appear in a correct synthesized program, and the
structure in which they are most likely to do so in. We use two probabilistic
models, \IID{} and \markov{} to do this (\bfsc{prediction}). The result of this
step is a distribution over program sketches.

The program synthesis phase (\bfsc{synthesis}) samples from this distribution to
obtain potential program sketches. These are then filled in with instructions to
form a candidate program. Candidates are then evaluated on the inputs to see if
they match the output. If one does, a correct solution is reported.

\subsection{Example}

In order to illustrate the workflow illustrated in \Cref{fig:system}, it is
worth considering each step in the context of an example function. We begin with
the specification:

\begin{minted}{C}
float f(float *a, float *b, int c) {
  // implementation details are unknown
}
\end{minted}

The first step is to sample random input values for this type signature, then
pass them to the function to observe the output value. For a single example:
\begin{minted}{C}
float a[] = { 0.0, 1.2, -3.4, -5.6 };
float b[] = { -1.0, 1.2, 2.4, 3.2 };
int c = 3;
float ret = f(a, b, c); // ret == -6.72
\end{minted}

Next, the most likely candidates from the population of sketch fragments are
identified. For this problem, a subset of the initial population resembles:
\[
  \{\; \mathrm{loop}(a), \mathrm{if}(c), \mathrm{loop}(), \dots,
  \mathrm{loop}(c), \mathrm{affine}(), \mathrm{index}() \;\}
\]

A full description of the semantics of each of these fragment types is given in
\Cref{sec:fragments}. From this population, our \IID{} model identifies the
fragments likely to appear in a solution. For this example type signature, these
are:
\[
  \{\; \mathrm{loop}(a,c), \mathrm{loop}(b,c), \mathrm{loop}(a,b,c),
  \mathrm{linear}() \;\}
\]

Next, our \markov{} model identifies the probability that a particular sequence
of these fragments will form the structure of a correct solution:
\begin{gather*}
  \mathbb{P}(\mathrm{loop}(a,b,c) \circ \mathrm{linear}()) = 0.2 \\
  \dots \\
  \mathbb{P}(\mathrm{linear}() \circ \mathrm{loop}(c)) = 0.01
\end{gather*}

Sketches are sampled using these probabilities. For each sketch, code is
generated. For the most likely composition above, as C:
\begin{minted}{C}
float f(float *a, float *b, int c) {
  for(int i = 0; i < c; ++i) {
    float ea = a[i], eb = b[i];
  }
}
\end{minted}

Finally, additional instructions are enumerated and added to the generated code
to produce solutions. In the case of this example, a correct program is:
\begin{minted}{C}
float f(float *a, float *b, int c) {
  float g = 0.0f;
  for(int i = 0; i < c; ++i) {
    float d = a[i], e = b[i];
    float h = d * e;
    g = g + h;
  }
  return g;
}
\end{minted}

By probabilistically identifying likely control flow structures, the synthesis
of a complex looping program is reduced to a smaller enumerative search.
% This
% section provides an illustrative example of the processes used by \presyn{};
\Cref{sec:fragments,sec:models,sec:synthesis} give detailed insight into each
of these steps individually.

\section{Specification} \label{sec:specification}
The primary input given to any program synthesizer is a \emph{specification}
describing the problem for which to synthesize a solution. While some
synthesizers provide specifications in the form of manually constructed
examples, \presyn{} does not: it specifies synthesis problems in terms of an
existing implementation; the goal is to capture the behavior with a synthesized
solution.

\presyn{} uses only two inputs to specify a synthesis problem: the type
signature of the target function, and a shared library containing an
implementation with that signature. There are no restrictions on the internal
details of this implementation.

% only that it has the correct signature and can
% be loaded dynamically.

\subsection{Signature}

\presyn{} aims for interoperability with C; the programs
it synthesizes should be ABI-compatible with the reference implementation. \presyn{} supports the
primitive C types \mintinline{C}{char}, \mintinline{C}{int} and
\mintinline{C}{float}, pointers to these types and \mintinline{C}{void *}.

% Extending the type system to the full C type system is not a conceptual change
% to \presyn{} (i.e.\ implementing it is a question of engineering work).

This type system allows for greater flexibility in specification than many other
synthesizers do. For example, a common restriction is for synthesizers to only
consider programs from lists to lists, or lists to single values
\cite{Feng2018,Balog2016}. In this model, programs are functional and cannot
modify state, a restriction not shared by \presyn{}.  By using an existing,
real-world type system, \presyn{} expresses problems more naturally than other
synthesizers are able to.

\subsection{IO Examples}

\presyn{} generates scalar input parameters by sampling uniformly from
fixed-size intervals. Pointer input data is generated by allocating a block of
memory and generating each element using the scalar generation strategy. Input
data is then passed to the target component; the return value and any pointer
parameters make up the recorded output.

We operate under the assumption that observational equivalence (i.e.\
equivalence over a large set of IO examples) is sufficient. This assumption is
shared by other work \cite{An2019, Claessen2000}. While in theory functions may
exhibit different behavior on a sparse subset of the input space, we did not
find any examples of this in practice.

\section{Fragments} \label{sec:fragments}
\presyn{} uses component-based, sketching program synthesis to construct
programs; the structure of a solutions is determined by the composition
of smaller parts, or \emph{fragments}.

\paragraph{Definition} We define a set $ \mathbf{F} $ of fragments, and a set $
\mathbf{P} $ containing concrete programs in a target language, with supported
operations:
\begin{align*}
	\mathit{compose} & : \mathbf{F} \times \mathbf{F} \to \mathbf{F} \\
	\mathit{compile} & : \mathbf{F} \to \mathbf{P}
\end{align*}
These operations are total; 
% any two fragments can be composed to yield another
% fragment, and any fragment can be compiled to a concrete program. 
there are no
invalid or intermediate states in this representation. For fragments $ a, b, c $
we write:
\begin{align*}
	a \circ b & \triangleq \mathit{compose}(a, b) \\
  a \circ b \circ c & \triangleq (a \circ b) \circ c
\end{align*}

Additionally, we consider functions onto $ \mathbf{F} $ (\emph{templates}). For
example, the function
\begin{align*}
	\mathit{fixed\mhyphen{}loop} : \mathbb{N} \to \mathbf{F}
\end{align*}
represents a fragment template for loops to a fixed upper bound $ \in \mathbb{N}
$; we use templates to parameterize fragments over argument names in a
specification.

\paragraph{Example} The semantics of $\mathit{compose}$ and $\mathit{compile}$
are defined by the implementation of each fragment in $ \mathbf{F} $. Consider
the fragments $ \mathit{fixed\mhyphen{}loop}(5) $ and $ \mathit{skip} $. The
definition of $ \mathit{compile} $ for these fragments (to C) is:
\begin{align*}
	\mathit{compile}(\mathit{fixed\mhyphen{}loop}(5)) & \triangleq \mintinline{C}{for(int
  i=0;i<5;++i) {}} \\
  \mathit{compile}(\mathit{skip}) & \triangleq \mintinline{C}{{}}
\end{align*}
The $ \mathit{fixed\mhyphen{}loop} $ fragment compiles to an empty loop, while
the $ \mathit{skip} $ compiles to an empty statement. Composition as defined by
each fragment produces more interesting structure. For all $ f \in
\mathbf{F} $:
\begin{align*}
	\mathit{compose}(\mathit{fixed\mhyphen{}loop}(5), f) \triangleq & \; \mathit{fixed\mhyphen{}loop}(5)_f \\
	\mathit{compile}(\mathit{fixed\mhyphen{}loop}(5)_f) \triangleq \\
	\mintinline{C}{for(int i=0;i<5;++i)} & \;\mathtt{\{} \mathit{compile}(f) \mathtt{\}}
\end{align*}

\subsection{Fragment Population}

\presyn{} uses a library of 11 different types of fragment; an overview of
these is given in \Cref{tab:fragments}.

\begin{table}
  \centering
  \caption{Summary of fragments used by \presyn{} to perform synthesis,
  organized by their high-level semantic categories.}
  \label{tab:fragments}
  \begin{tabularx}{\columnwidth}{l|rX}
    \toprule
      \textbf{Group} & \textbf{N} & \textbf{Description}  \\
    \midrule
      Computation & 5 & Control flow-free regions of code that represent ``holes''
      into which concrete instructions can be instantiated. Includes specialized
      structures (e.g.\ affine array index expressions). \\
    \midrule
      Iteration & 3 & Loop variants with different conditions for termination
      (e.g.\ a fixed upper bound, or a predicate test). \\
    \midrule
      Control Flow & 3 & Conditional control flow, as well as sequential
      execution of sets of fragments. \\
    \bottomrule
  \end{tabularx}
\end{table}

\begin{figure*}
  \centering
  \includegraphics[width=\textwidth]{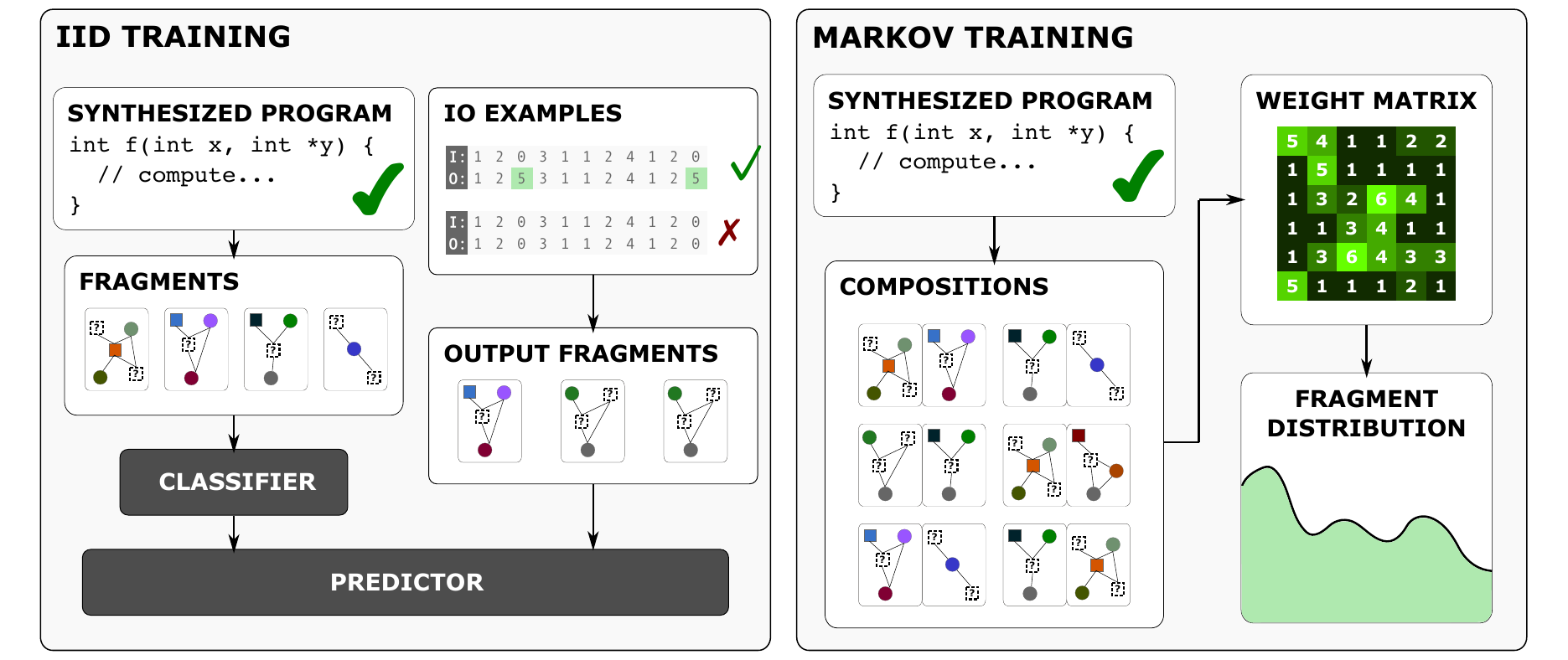}
  \caption{
    Training process for our \IID{} and \markov{} models based on corpus data.
  }
  \label{fig:training}
\end{figure*}

% \begin{description}
%   \item[Linear] A basic block into which instructions should later be
%     synthesized; an empty variant prevents instructions.

%   \item[Fixed Loop] Template parameterized on an optional pointer
%     \mintinline{C}{ptr} and an integer \mintinline{C}{x}:
% \begin{minted}{C}
% for(int i = 0; i < x; ++i) { ? }
% for(int i = 0; i < x; ++i) { use(ptr[i]); ? }
% \end{minted}

%   \item[Delimiter Loop] Template parameterized on a pointer \mintinline{C}{ptr}:
% \begin{minted}{C}
% while(*ptr++ != _P) { use(*ptr); ? }
% \end{minted}

%   \item[Loop] A catch-all for iterations not covered by the two specialized
%     types:
% \begin{minted}{C}
% while(_P) { ? }
% \end{minted}

%   \item[If and If-Else] Conditional control flow:
% \begin{minted}{C}
% if(_P) { ? }
% if(_P) { ? } else { ? }
% \end{minted}

%   \item[Seq] Execute two fragments, one after the other:
% \begin{minted}{C}
% ? ; ?
% \end{minted}

%   \item[Affine and Index] Synthesize affine and general index expressions
%     respectively, parameterized on \mintinline{C}{ptr}. For example:
% \begin{minted}{C}
% int a_v = ptr[_P * _P + _P]; // e.g.
% int v = ptr[_P - _P];        // e.g.
% \end{minted}

% \end{description}

The aim for these components was to remain as general as possible so that
\presyn{} can express many different programs, while not biasing towards one
problem domain at the expense of others. We selected some fragments based on
intuition for common programming practices, and the remainder from high-level
idiomatic patterns used in compiler analyses \cite{Ginsbach2018}.

\section{Probabilistic Models} \label{sec:models}
\presyn{} uses two probabilistic models to guide synthesis. The first, \IID{},
predicts whether each fragment in the population is likely to appear in a
solution or not. The second, \markov{}, creates a probability distribution over
sequences of fragments that form compositions.

\subsection{\IID{}}

After generating input-output examples (\bfsc{specification} in
\Cref{fig:system}), \presyn{} collects an initial set of fragments $
\mathbf{F_0} \subseteq \mathbf{F} $ from which solutions may be composed
(\bfsc{prediction} in \Cref{fig:system}). The library of fragment templates used
by \presyn{} is too large to perform an exhaustive search, and so accurately
\emph{predicting} likely fragments is important.

Suppose $ f_0 , f_1 , \dots , f_n $ is an ordered sequence of fragments that
when composed, produces a correct solution for a synthesis problem. We aim to
predict an initial set of fragments $ \mathbf{F_0} $ that matches 
$ \{ f_0 , f_1 , \dots , f_n \} $ as accurately as possible (without considering
ordering). To do so, we combine two ideas:

\paragraph{Fragment Semantics} It is possible to make observations about the
behavior of the reference implementation based on the input-output examples
generated for testing. For example, if a memory region is written to in any of
the observed examples, then that region represents an output reference
parameter. If such parameters are present, then the set of fragments capable
of performing output is collected as $ \mathbf{F_S} $.

\paragraph{Classification Model} As well as semantic knowledge, we employ a
simple classification model to determine inclusion in $ \mathbf{F_0} $ for
non-output fragments. To do this, we require a small corpus of training data:
type signatures and $ \mathbf{F_0} $ for successfully synthesized
programs. We train a random forest model, using the type signatures as the input
and a binary inclusion indicator for each fragment as the target. This trained
model provides a decision function $ P $ that determines fragment inclusion in $
\mathbf{F_0} $.

Writing $ \mathbf{F_P} $ for the set of fragments satisfying the trained decision
function $ P $, we define:
\[
  \mathbf{F_0} \triangleq \mathbf{F_S} \, \cup \; \mathbf{F_P}
\]

It is clear that it is safe for the prediction of $ \mathbf{F_0} $ to
over-approximate the true initial set: if additional irrelevant fragments are
present, synthesis will simply be slower. However, $ \mathbf{F_0} $ may in fact
be an \emph{under}-approximation. This is addressed by the next model in the
synthesis process, \markov{}.

\subsection{\markov{}}

The set of initial fragments $ \mathbf{F_0} $ represents the starting point for
synthesis. However, in order to produce a solution structure, fragments must be
composed in the correct order. Additionally, the process of generating these
compositions should be robust if $ \mathbf{F_0} $ is an under-approximation.

Our model for generating these compositions uses the same training data as
\IID{}, but treats fragment occurrences as a Markov model rather than a
collection of IID random variables. This model is defined as follows:

\paragraph{Definitions} We define auxiliary fragments $ f_{\mathit{start}},
f_{\mathit{end}} $ that both act as the identity under composition, along with a
function $ w $ such that $ w(f, f') $ represents the number of occurrences of
the composition $ f \circ f' $ in the training set.

\paragraph{Training} As with \IID{}, training data for \markov{}
comprises correctly synthesized programs together with the composition of
fragments used to generate that program. To train the model, each observed
composition of fragments has $ f_{\mathit{start}} $ prepended and $
f_{\mathit{end}} $ appended. The composition sequences are then split into
pairs, and the number of occurrences of each unique pair is recorded and
used to define $ w $.

\paragraph{Generating a Composition} Sequences of fragments (representing
compositions) can be sampled from \markov{} as follows. First, we define:
\begin{align*} 
  s(f) & \triangleq \sum_{f'\in \mathbf{F}} w(f, f') 
\end{align*}
Then, starting from $ f_{\mathit{start}} $ we sample fragments with conditional
probability:
\[
  \mathbb{P}_w(f_n | f_{n-1}) = \frac{w(f_{n-1}, f_n)}{s(f_{n-1})}
\]

To include the predicted initial set of fragments $ \mathbf{F_0} $, we then
define an augmented model $ w' $:
\begin{align*}
  w'(f_i, f_j) & \triangleq w(f_i, f_j) \text{ if } f_j \in \mathbf{F_0} \\
               & \triangleq 0 \text{ else}
\end{align*}

The sum $ s' $ is defined identically to $ s $, but with respect to $ w' $
rather than $ w $. Similarly, $ \mathbb{P}'_{w} $ is defined with respect to $
w' $ and $ s' $ rather than $ w $ and $ s $. With a weighting parameter $ b \in
[0, 1] $ chosen, fragments are sampled with probability:
\[
  \mathbb{P}(f_n | f_{n-1}) = b\mathbb{P}'_w(f_n | f_{n-1}) + (1 - b)\mathbb{P}_w(f_n | f_{n-1})
\]

A sequence of fragments $ f_0 \circ \dots \circ f_n $ is generated by sampling
from this distribution until $ f_{end} $ is sampled, or to a fixed maximum
length. The sequence is then composed (from left to right) to produce a program
sketch.

\section{Synthesis} \label{sec:synthesis}
The next stage in the synthesis process is to generate concrete programs from
candidate sketches generated by our \markov{} model. They are then executed and
tested against the input-output examples collected at the specification step.

Fragments can be compiled to a concrete program in LLVM \cite{Lattner2002}
intermediate representation, but the resulting programs
cannot yet be executed:
\begin{itemize} 
  \item They do not perform any meaningful computation; the programs do not yet
    contain any instructions beyond those used for control flow and program
    structure.

  \item The programs contain \emph{placeholders}: typed SSA variables
    in the program that do not yet have a defined value. For example, a fragment
    that implements a while-loop may use a boolean placeholder as the loop
    condition. It cannot be executed until a concrete value is chosen to fill
    the placeholder.
\end{itemize}

To resolve these issues, we traverse the dominance tree of the compiled function
in-order. For each basic block, we maintain a set of instructions that could be
added to it based on the live SSA values at the block's entry. We use an
enumerative search to instantiate different concrete programs from the compiled
fragment. Once a set of instructions has been selected by this search, a value
for each $\phi$ node and placeholder is chosen by a secondary search.

\paragraph{Safety} Because synthesized programs are able to perform potentially
arbitrary memory accesses, \presyn{} implements a conservative, compile-time
bounds checking system. During synthesis, all allocated memory passed to the
synthesized programs has the same size.  If programs make out-of-bounds
accesses, the size is increased for future attempts up to a configured maximum.
Bounds-checking code is removed when reporting solutions to accurately reflect
algorithmic intent.

\subsection{Testing}

\presyn{} specifies problems by collecting a large number of input-output
examples from the black-box component. The correctness of a candidate program is
specified as \emph{observational equivalence}: if the candidate program behaves
identically to the reference component on every input generated, then it is
correct.
% Floating point data is tested using an approximate notion of equality; either
% ULP distance or absolute difference with configurable thresholds.

While it is possible that observational equality (even over a large set of
examples) is unsound, it is not possible to implement a better decision
procedure for an arbitrary black box. This is an observation shared by other
work \cite{An2019}.

\section{Experimental Setup} \label{sec:setup}
Benchmarking program synthesizers against each other fairly while allowing for
differences in specification and expression is a challenging problem
\cite{Pantridge2017,Helmuth2015}. This section describes our experimental
methodology for evaluating \presyn{} and comparing its performance to other
program synthesizers.

\subsection{Overview of Methodology}

%The aim of our evaluation is to determine whether \presyn{} can effectively
%capture the computational structure of black-box components by using only their
%observed behavior. 
We identify a collection of benchmark synthesis problems
collated from existing work on program synthesis, as well as from real-world
software components. For each of these problems, we prepare a specification for
\presyn{} as described in \Cref{sec:specification}, then attempt to synthesize a
solution.

%As well as a straightforward evaluation of \presyn{} on benchmark problems.
We evaluated  \presyn{} against 
five other state-of-the art program
synthesizers.
% (\baseline{} \cite{Collie2019}, \sketchadapt{} \cite{Nye2019}, \makespeare{} \cite{Rosin2018},
%\simpl{} \cite{So2017} and \LL{} \cite{Feser2015}) on the same set of problems.
Doing so requires benchmarks to be fairly translated;
%into the specification format for each synthesizer;
a non-trivial task when accounting for the
differences between them.
%MOB
%In particular, some of the synthesizers needed additional help beyond IO examples, while others %required
%significantly more time  to successfully synthesize.

\begin{comment}
Finally, we investigate the accuracy and impact of \presyn{}'s internal models.
The prediction accuracy of $ \mathbf{F_0} $ is measured using the Jaccard
coefficient for each synthesis problem.  Then, we compare synthesis performance
in two scenarios: using only our \IID{} model, and using the \markov{} model as
well.
\end{comment}

\subsection{Implementations}

\begin{comment}
\presyn{} performs synthesis based on the behavior of a black box component
using observations of that component's behavior. This style of synthesis is less
common than human-specified, example-driven synthesis.  We identified one other
synthesizer based on black box specifications, and three that use handwritten
examples but that could be fairly compared to \presyn{}.
\end{comment}

We had four primary criteria for selecting implementations to compare against.
First, they should target a `general-purpose' language. This ruled out
synthesizers such as 
%\textsc{Cosette}
 \cite{Chu2017} or
%\textsc{Nonograms} 
\cite{Butler2017} that target specific problem domains using
SMT-aided formal techniques.  Typically, the problem space for programs with
arbitrary control flow is too large for these solver-aided techniques
\cite{So2017} and other methods are required. Second, the implementations
selected should be representative of the current baseline synthesis performance
for target languages similar to their own. Third, the set of implementations
should cover a range of different program synthesis methods. Finally, the
implementations should be available for evaluation and testing; we aim to make
our own implementation and evaluation available in turn.
% through the artifact
% evaluation process.
With these goals in mind, the implementations selected are:

\paragraph{\sketchadapt{}} \cite{Nye2019} is a neural synthesis approach. %and most similar in 
%spirit to \presyn{}. 
It samples programs from a DSL to generate training data for a generative model
that predicts programs from IO examples. Its chief innovation is to adapt to IO
complexity, falling back to synthesis when generation is predicted to be too
expensive. It represents the state-of-the-art in neural synthesis.

\paragraph{\baseline{}} \cite{Collie2019} uses program synthesis to enable
refactoring of legacy scientific software to use new libraries.  It uses a
black-box specification to generate IO examples similarly to \presyn{}, but requires
extensive annotation of function signatures to do so.

\paragraph{\makespeare{}} \cite{Rosin2018} represents the state-of-the-art for
genetic program synthesis \cite{Banzhaf1998} of programs with complex control
flow, and contributes a novel hill-climbing algorithm. Synthesized programs are
in a subset of \textsc{x86} assembly, and problems are specified by providing
`before and after' memory and register state for an abstract machine.

\paragraph{\simpl{}} \cite{So2017} targets imperative programs written in a
small C-like language. Its key innovation is the use of static analysis
techniques to prune the search space of candidate programs to ensure that `dead
ends' are not explored unnecessarily. The programs it is evaluated on are
designed to model introductory programming exercises with control flow and
mutable variables; problems are specified using handwritten input-output
examples and a partial program sketch.

\paragraph{\LL{}} \cite{Feser2015} targets a functional language that permits
algebraic data types. 
%It aims to provide a provably
%minimal and general solution to data-structure manipulation problems, by
It  uses  a
type-aware recursive search process over a space of expressions. Programs
synthesized are compositions of a standard library of functions 
%(e.g.\integer manipulations
 and higher-order functional primitives. 
% such as map and filter)
 Like \simpl{}, problems are specified by handwritten input-output
examples, and in some cases a recursive base case must be supplied.

\subsection{Benchmark Collation}

\begin{figure*}
  \centering

  \captionof{table}{Groups of synthesis benchmark problems}
  \label{tab:problems}

  \tabulinesep=1mm
  \begin{tabu} to \textwidth {rrX}
    \toprule
    \rowfont{\bfseries} Group & N & Description \\
    \midrule

    makespeare & 11 &
    Problems that require loops to manipulate arrays of integers in place. We
    use the full set of benchmarks from \cite{Rosin2018}, modulo those adapted
    from \cite{So2017}.
    \\

    simpl-int & 15 &
    Arithmetic manipulations of integer values, requiring loops and
    data-dependent control flow. We use the full set of integer benchmarks from
    \cite{So2017}.
    \\

    simpl-array & 12 &
    Problems stated over integer arrays, with different styles required (e.g.\
    pairwise iteration, reductions and elementwise computation). We use the full
    set of array benchmarks from \cite{So2017}.
    \\

    $\lambda^2$ & 8 &
    Singly-nested integer linked-list manipulation problems from
    \cite{Feser2015}, restated for other synthesizers as array problems.
    \\

	SketchAdapt & 10 &
	A series of generated list problems, taken as representative samples
	from the 500 program evaluation file presented by \cite{Nye2019}.
	\\

    \midrule

    string & 16 &
    The C standard library's string processing functions \cite{stringh}. We
    remove impure functions such as \texttt{strtok}.
    \\

    mathfu & 15 &
    Vector-scalar and vector-vector mathematical functions from the Mathfu
    \cite{Mathfu} library.
    \\

    blas & 4 &
    Matrix-vector linear algebra functions from the BLAS \cite{Blackford2001}
    standard as synthesized in \cite{Collie2019}. We disregard functions such as
    \texttt{spmv} for which extensive assistance was required.
    \\

    dsp & 21 &
    Vector- and matrix-based signal processing functions from the TI signal
    processing library \cite{TIDSP}, adapted for platform portability and
    removing functions with requirements for extensive constant data such as
    filter taps.
    \\
    \bottomrule
  \end{tabu}
\end{figure*}

We collated a set of 112 synthesis problems from two sources: the benchmarks
used by existing synthesizers, and real-world software libraries. The sets of
problems selected are summarized in \Cref{tab:problems}.

\paragraph{Existing Benchmarks} Each synthesizer evaluated uses its own set of
synthesis benchmarks, with some partially overlapping. We began with the full
set of benchmarks from each of \cite{Rosin2018,So2017,Feser2015} then removed
duplicate entries. We then added a representative sample of problems from
\cite{Nye2019}. To create a more level playing field, we then removed
problems for which only one synthesizer was specialized towards. For example,
\baseline{} includes extensive type annotations that allow it to synthesize
sparse-matrix vector multiplication, and \LL{} includes tree-processing
primitive functions.

\paragraph{Black-Box Components} In \cite{Collie2019}, \baseline{} is used to
synthesize implementations of 11 functions shared between a set of optimized
mathematical libraries such as Intel MKL \cite{mkl} or Nvidia cuBLAS
\cite{cublas}. We extended this set to include additional mathematical
operations found in other real-world libraries: Mathfu \cite{Mathfu} and the TI
signal processing libraries \cite{TIDSP}.  Additionally, we identified string
processing functions as a common target domain for program synthesis
\cite{Gulwani2011,Perelman2014,Parisotto2016a}. 
%Typically these programs are
%composed from transformations in a DSL with very expressive primitives. In order
%to allow for each synthesizer to express the programs, 
We identified the C
standard library string functions \cite{stringh} as a realistic target.

\subsection{Problem Preparation}

The synthesis benchmarks we selected are all specified using different formats.
%and so inputs need to be prepared differently for each one. Our methodology was
%as follows:
For each problem, we therefore produced a reference implementation in C based on the
specifications in the original papers for existing benchmarks, and based on the
concrete implementation for real-world code. 
We then generated an appropriate
number of IO examples in the correct format for each one.

In some cases, this required some adaptation.
For example, most synthesizers do not support floating point computation; we
restated these using integers where appropriate, with the aim of preserving the
intent of each problem.

\subsection{Synthesizer Help}
\label{help}
Although we are interested in black-box IO example synthesis, many of the synthesizers 
rely on varying degrees of help. We evaluate them  with and without the following assistance:

\paragraph{\sketchadapt{}} did not receive any extra help to solve each
	problem.  It does has some difficulties surrounding
	the integer type, so we convert integers to singleton lists
	where invalid input types are detected.
\paragraph{\baseline{}}
        \baseline{} requires semantic annotations to be applied to the type
        signature of a target function (for example, ``the pointer \texttt{x}
        points to \texttt{N} elements''). These annotations are used as
        heuristics to guide the search for potential candidate structures.
	
\paragraph{\makespeare{}} uses  a small number of registers  values 
        which provide some aid to guiding program generation, however in practice this had little impact on behavior.
        Unlike other schemes it required a large number of
		examples, typically in the thousands.  We observe that \makespeare{}
		is dependent on the large, and varied, set of input examples
		in many cases.

\paragraph{\simpl{}} relies on  a partial program and
	a list of useful integer constants  as input.
	We provide  the correct number
	of top-level loops and the correct number of variable initializations,
	which is consistent with many (but not all) of their benchmarks.
\paragraph{\LL{}} can exploit an extensible library of base cases.
        We provide accesses to its standard library of base cases 
\paragraph{\presyn{}} does not require additional help.

\section{Results} \label{sec:results}
\begin{figure*}
  \centering
  \includegraphics[width=\textwidth]{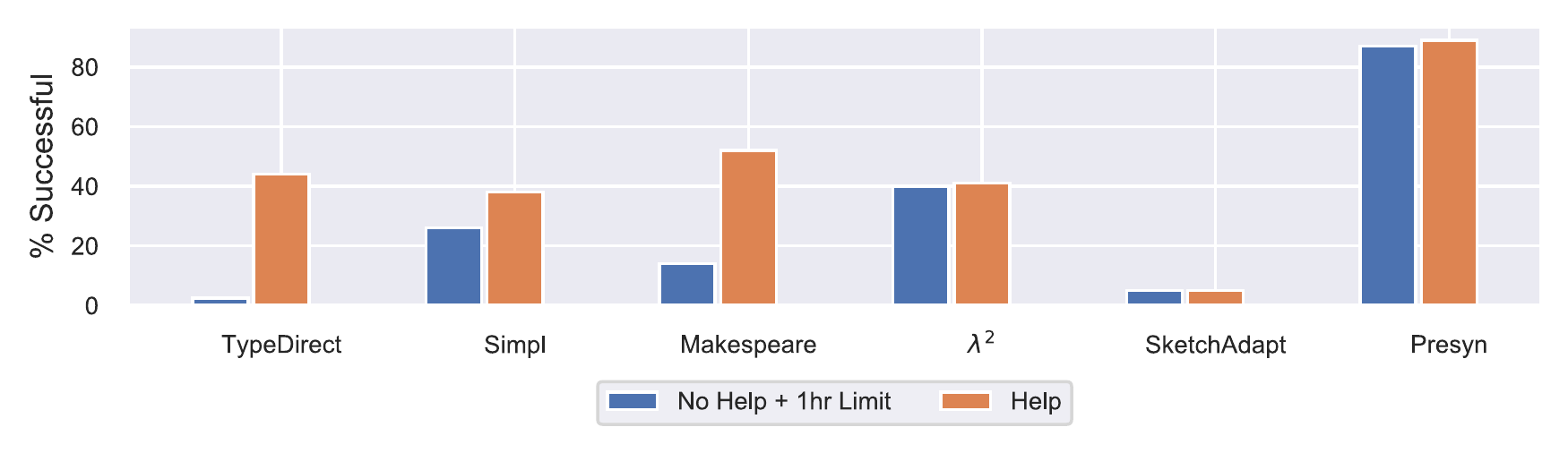}
  \caption{
    Proportion of the synthesis benchmark set synthesized by each implementation
    under favorable conditions (see section \ref{help}), and when restricted by time limits and reduced
    help.
  }
  \label{fig:coverage}
\end{figure*}

\begin{table*}
  \centering
  
  \caption{
    Proportion of each group of synthesis benchmarks synthesized by each
    synthesizer under favorable conditions (see section \ref{help})}
  \label{tab:perf}

  \begin{tabularx}{\textwidth}{lr|YYccY|Y}
    \toprule
        & \textbf{Group} & \textbf{\baseline{}} & \textbf{\makespeare{}} 
        & \textbf{\simpl{}} & \textbf{\LL{}} & \textbf{\sketchadapt{}}
        & \textbf{\presyn{}} \\

    \midrule
      \multirow{5}{*}{Benchmarks}
      & makespeare    & 0.20 & 0.64 & 0.09 & 0.45 & 0.00 & 0.55 \\
      & simpl-int     & 0.00 & 0.80 & 0.73 & 0.20 & 0.00 & 0.93 \\
      & simpl-array   & 0.58 & 0.75 & 0.58 & 0.58 & 0.08 & 0.92 \\
      & $\lambda^2$   & 0.43 & 1.00 & 0.38 & 0.75 & 0.00 & 1.00 \\
      & SketchAdapt   & 0.20 & 0.50 & 0.00 & 0.10 & 0.10 & 0.50 \\
      & \textbf{Mean}
          & \textbf{0.26} & \textbf{0.73} & \textbf{0.39} & \textbf{0.39} & \textbf{0.04} & \textbf{0.79} \\
    
    \midrule
    \multirow{5}{*}{Libraries}
      & string        & 0.00 & 0.23 & 0.13 & 0.56 & 0.00 & 0.75 \\
      & mathfu        & 1.00 & 0.60 & 0.67 & 0.47 & 0.13 & 1.00 \\
      & blas          & 0.75 & 0.00 & 0.00 & 0.00 & 0.00 & 1.00 \\
      & dsp           & 0.90 & 0.26 & 0.40 & 0.38 & 0.00 & 1.00 \\

        & \textbf{Mean}
        & \textbf{0.92} & \textbf{0.33} & \textbf{0.38} & \textbf{0.48} &
    \textbf{0.04} & \textbf{0.93} \\
        \midrule
        & \textbf{Mean}
        & \textbf{0.53} & \textbf{0.54} & \textbf{0.39} & \textbf{0.43} &
    \textbf{0.04} & \textbf{0.86} \\
    \bottomrule
  \end{tabularx}

\end{table*}

This section presents our experimental results based on the framework presented
in the previous section. For cross-evaluation and when comparing \IID{} to
\markov{}, \presyn{} was trained on a randomly selected subset (15\%) of the
synthesis problems for which we identified correct fragment structure.

\subsection{Coverage}

The primary evaluation criteria for \presyn{} against other synthesizers is the
number of programs it is able to synthesize. We evaluated each synthesizer in
three contexts for each synthesis problem: with no additional help beyond IO
examples for the problem (i.e.\ no annotated types or program structure), with
the appropriate help as described in \Cref{help} and a conservative timeout, and finally unrestricted (with
help). These results are shown in \Cref{fig:coverage}.

\presyn{} is able to successfully synthesize more functions across the set of
synthesis benchmarks than each of the other implementations, even when they are
given appropriate assistance and unlimited execution time: 89\% of the functions
evaluated, while the next-best performing (\makespeare{}) synthesizes only 65\%.
On real-world code, \presyn{} synthesizes 93\% vs.\ 63\% for \baseline{}. The
full results for each synthesizer and library are given in \Cref{tab:perf}.
Interestingly, it is not the case that each synthesizer performs best on its own
benchmarks or that each set of benchmarks is best synthesized with the
corresponding implementation; this is likely due to the differences in setup
between our experiments and the original work. Nonetheless, it indicates that
synthesis is by nature a fragile problem to evaluate experimentally.

\makespeare{} required a large number of
	examples, typically in the thousands but  for some
	programs, e.g.\ factorial, thousands of examples would not
	fit within normal-width integers.
	\makespeare{} did not seem particularly influenced by
	help but requires an extremely long time to synthesize 

For \simpl{} we found that ignoring the partial program input entirely produced
	very poor results. It struggled significantly on many
	benchmarks without the desired program structure as input.

For \LL{}, there were a small number of examples where providing a recursive
base case for a problem as help made a difference.

\sketchadapt{} suffered from  poor performance on general-purpose
problems.  We found that SketchAdapt was only successful on trivial
examples outside its own evaluation domain and in fact found that
it was unable to reproduce its own results once the input values were
changed. As it has the least mature implementation, this is perhaps not unexpected.
Further, we have taken \sketchadapt{} further out of its comfort zone
than any of the other synthesizers evaluated here---it is intended to
perform well on restricted problem domains where the
generalizations required are much smaller where the dataset covers
a more significant fraction of the program space.  Examples of
this can be seen within the original paper, where \citet{Nye2019}
find high accuracy on a number of different subdomains.

When synthesis time is limited or less help is provided for a synthesis problem,
\presyn{} exhibits an even greater advantage over other implementations. Both
\LL{} and \simpl{} exhibit degraded synthesis when assistance is not given (not
shown in \Cref{fig:coverage} is that successful syntheses took up to $ 300
\times $ longer to discover in these cases). \makespeare{}'s use of a genetic
algorithm means that it relies on being able to spend a long time searching a
space of programs, and struggles when a timeout is imposed.

\subsection{Synthesis time and validity}

The amount of time spent in synthesis by each scheme varied considerably.
\presyn{}, \LL{} and \simpl{} all showed mean synthesis times (for successful
cases) of less than \SI{120}{s}. \sketchadapt{} required longer, with a mean
synthesis time of \SI{914}{s}. Because of its reliance on genetic search,
\makespeare{} used a mean of \SI{4 522}{s} per synthesized program, with some
taking up to 3$\times$ this long before being timed out.

The size of programs generated by \presyn{} varied from 40 to 110 lines of LLVM
IR. As we do not have a formal specification, we can only test synthesized
programs, not verify them. For every \presyn{} synthesized program, we
automatically generated random and boundary value inputs and checked if outputs
matched those from the target black-box function. In all cases we find them
behaviorally equivalent.

Given that we have access to the component code generating the IO examples, we
manually inspected the synthesized results. In all cases, using our knowledge,
we judged them to be correct. Future work will examine the use of bounded model
checking as a means of providing greater assurance.

\subsection{Impact of Probabilistic Models}

\begin{table}
  \centering
  \caption{
    Jaccard coefficient for predictions of the initial fragment set $
    \mathbf{F_0} $ across each set of synthesis problems.
  }
  \label{tab:jaccard}

  \begin{tabu} to 0.7\columnwidth {lr|X[c]}
    \toprule
    \rowfont{\bfseries} & Group & Jaccard \\
    \midrule

    \multirow{4}{*}{Benchmarks} 
    & makespeare  & 0.77 \\
    & simpl-int   & 0.85 \\
    & simpl-array & 0.87 \\
    & $\lambda^2$ & 0.72 \\
    & SketchAdapt & 0.72 \\

    \midrule

    \multirow{4}{*}{Libraries} 
    & string      & 0.72 \\
    & mathfu      & 0.90 \\
    & blas        & 0.95 \\
    & dsp         & 0.77 \\

    \midrule

    \rowfont{\bfseries} \normalfont{All} & Mean & 0.81 \\

    \bottomrule
  \end{tabu}
\end{table}

\begin{figure}
  \centering
  \includegraphics[width=\columnwidth]{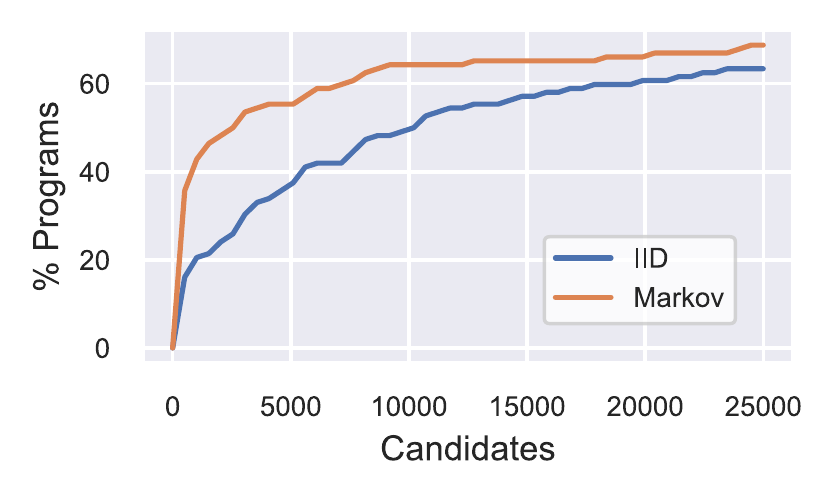}
  \caption{Functions synthesized vs.\ candidates evaluated using the IID and
  Markov models of synthesis.}
  \label{fig:progress}
\end{figure}

\paragraph{\IID{}} \presyn{} uses two probabilistic models to drive its
synthesis. The first, \IID{}, predicts fragments that are likely to form part of
a correct solution using a random forest model and a limited model of fragment
semantics. 

The accuracy of these predictions (measured using the Jaccard coefficient of
predicted and true $ \mathbf{F_0} $) is given in \Cref{tab:jaccard}. On average,
there is an 81\% overlap between the predicted value and the true value. \IID{}
is not significantly over- or under-approximating in its predictions; this
overlap corresponds to $ < 1 $ prediction errors per problem.

\paragraph{\markov{}} The second model aims to predict the correct order in
which to compose fragments to produce a program sketch. To evaluate \markov{},
we compare the number of programs synthesized by \presyn{} against only the
\IID{} model. These results are shown in \Cref{fig:progress}. We see that
\markov{} significantly accelerates the synthesis process; it is able to
synthesize 60 programs using fewer than half as many candidate programs. In the
`long tail' of programs, the difference is smaller as the complexity of
synthesis is dominated by the search for long sequences of instructions.

\subsection{Insights into Program Structure}

\begin{figure*}
  \centering
  \begin{subfigure}{\textwidth}
    \centering
    \includegraphics[width=\textwidth]{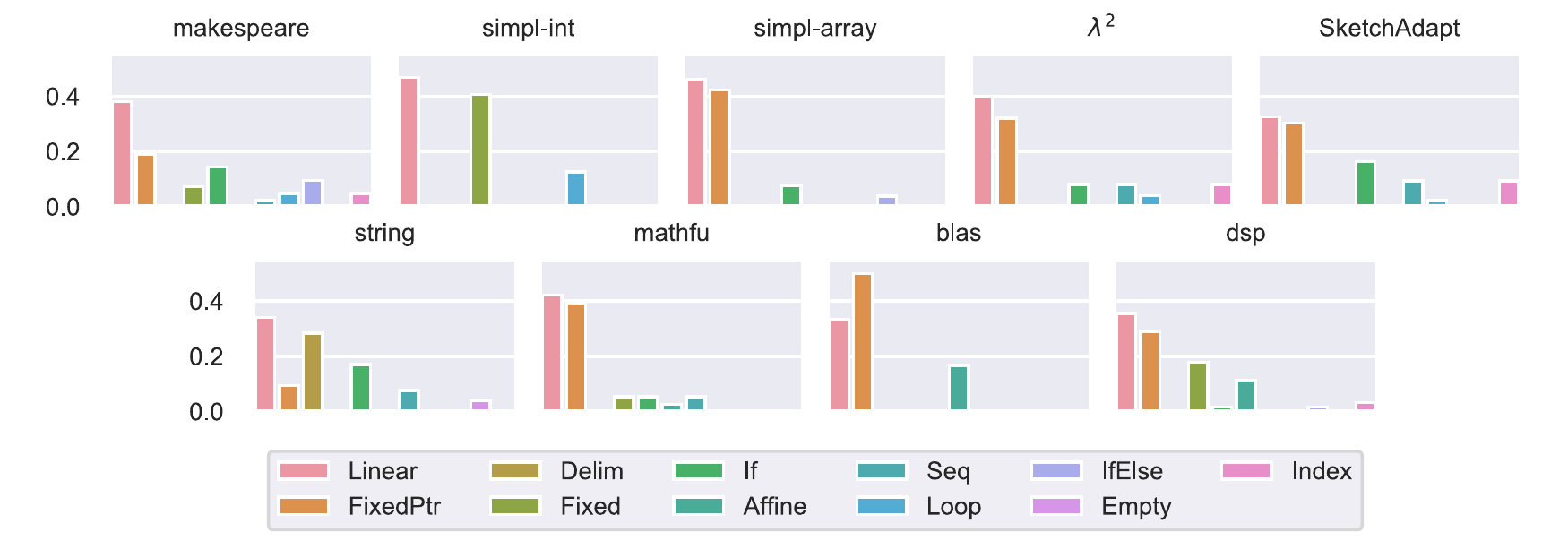}
    \caption{Relative frequency of each fragment type in each group of
    benchmarks.}
    \label{fig:freqs}
  \end{subfigure}
  \begin{subfigure}{\columnwidth}
    \centering
    \includegraphics[width=\textwidth]{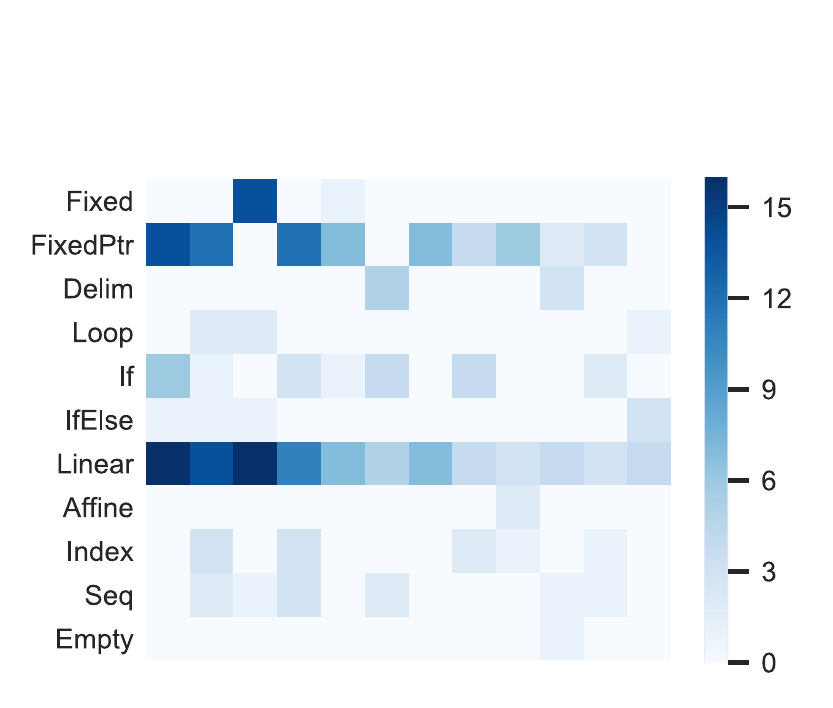}
    \caption{\IID{} distribution of each fragment type, organized by unique type
    signatures.}
    \label{fig:sigs}
  \end{subfigure}\hfill
  \begin{subfigure}{\columnwidth}
    \centering
    \includegraphics[width=\textwidth]{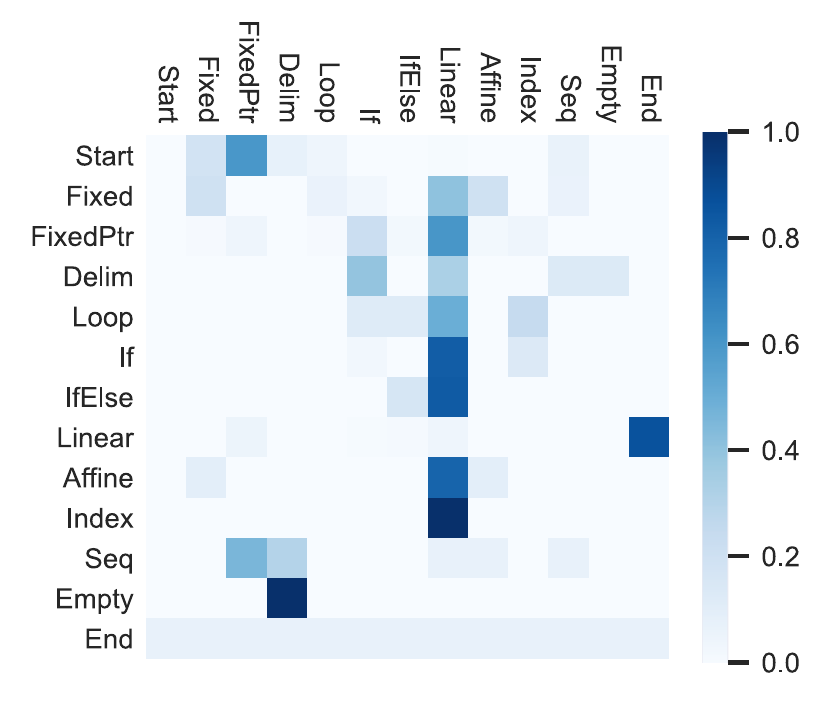}
    \caption{\markov{} model transition probabilities for each pair of
    fragment types: from-to is read along rows.}
    \label{fig:bigrams}
  \end{subfigure}

  \caption{Insights into fragment distributions produced by our \IID{} and
  \markov{} models.}
  \label{fig:insights}
\end{figure*}

As well as outperforming competing implementations on a wide range of synthesis
problems, \presyn{} provides interesting statistical insights into the structure
of the programs it synthesizes through its use of probabilistic models. The
results in this section summarize the distributions learned by \presyn{} over
its full set of synthesized programs: the models learned on the 15\% subset are
updated with all subsequent observations, but these are not used during
synthesis.

In \Cref{fig:insights} we show three different insights into the structure of
programs synthesized by \presyn{}. First, in \Cref{fig:freqs} we see the
relative frequency of each type of fragment across benchmark groups.  Linear
blocks of code are common across all the benchmarks; every program performs some
kind of computation. In terms of control flow, the easiest synthesis benchmark
suites (simpl-int, simpl-array, mathfu, \LL{}) are those with largely
homogeneous control flow across their benchmarks, while the more challenging
ones (makespeare, string, SketchAdapt) have far more variation. These results suggest that at
least an approximate notion of difficulty for a set of synthesis benchmarks is
the heterogeneity in code structure required to solve the problems in that set.
Other intuitive structure that can be observed is the ubiquity of nested loops
in the blas matrix-vector problems.

\Cref{fig:sigs} shows the number of fragments of each type that appear in
synthesized solutions, grouped by unique type signatures (for the 12 most common
signatures). Two patterns become apparent from this visualization: the most
common type signatures dominate the set of benchmarks, and the fragments used by
solutions are generally consistent for each type signature.

Finally, \Cref{fig:bigrams} shows the transition probabilities from our
\markov{} model. From this visualization, we can see that the model favors
generating smaller programs: most fragments are most likely to be followed by a
linear fragment, which is itself likely to end the synthesis.

\subsection{Case Studies}

Synthesizing programs that match the behavior of a black box has useful
applications in a software engineering context not enabled by other types of
synthesis; in this section we detail two of these use cases.

\subsubsection{Rejuvenating Legacy Code}

In \cite{Collie2019}, a system to improve the performance of legacy scientific
applications is proposed. It uses \baseline{} to synthesize programs matching
the behavior of library functions, such that compatible user code can be
remapped to use improved or optimized versions of the functions.

This approach led to considerable performance improvements of up to $ 10\times $
on deep learning models, and close to $ 2\times $ on real-world chemical
simulation benchmarks. However, the synthesis techniques used are limited: only
a small number of functions are considered, and annotations provided by the user
provide significant bias to the synthesis process.

As demonstrated by our results in \Cref{sec:results}, \presyn{} outperforms
\baseline{} across all the synthesis benchmarks we considered, while requiring
no annotations or assistance to be provided by the user for synthesis problems.
Of the 11 functions synthesized in \cite{Collie2019}, \presyn{} is able to
synthesize 10 (the exception being sparse matrix-vector multiplication, which
\baseline{} requires extensive assistance for). Additionally, we were able to
synthesize a further 12 functions from the libraries used in \cite{Collie2019}
using \presyn{}; all 22 syntheses were performed without manual type annotation.
By using \presyn{}, more opportunities for performance improvement and an
improved API migration can be generated.

\subsubsection{Detecting Library Reimplementations}

\newcommand{\pc}[1]{\pgfmathparse{#1}\pgfmathprintnumber{\pgfmathresult}\%}
\begin{table}
  \caption{
    Number of discovered instances in user applications where code duplicates
    library functionality.
  }
  \label{tab:match}
  \begin{tabu} to 0.8\columnwidth {lX[l]lr}
    \toprule
    \rowfont{\bfseries} Software & LoC & Library & N \\
    \midrule

    \multirow{3}{*}{FFmpeg} &
    \multirow{3}{*}{1,061,655}
    & Mathfu     & 10  \\
    && BLAS       & 3   \\
    && DSP        & 11  \\
    \midrule

    \multirow{1}{*}{Coreutils} &
    66,355
    & String     & 10  \\
    \midrule

    \multirow{2}{*}{GraphicsGems} &
    \multirow{2}{*}{46,619}
    & Mathfu       & 35 \\
    && DSP          & 26 \\
    \midrule

    \multirow{3}{*}{Darknet} &
    \multirow{3}{*}{21,299}
    & BLAS         & 5 \\
    && Mathfu       & 3 \\
    && DSP          & 5 \\
    \midrule

    \multirow{1}{*}{Nanvix} &
    \multirow{1}{*}{11,226} 
    & String       & 10 \\
    \midrule

    \multirow{2}{*}{ETR} &
    \multirow{2}{*}{2,399}
    & Mathfu       & 29 \\
    && DSP          & 16 \\
    \midrule

    \rowfont{\bfseries} Total && & 163 \\
    \bottomrule
  \end{tabu}
\end{table}

A well-known problem in software engineering is code duplication, particularly
when the duplicated functionality has already been implemented by a third-party
library \cite{Roy2014}. We performed an initial case study to show the
application of \presyn{} to this problem, discovering 163 instances in
real-world applications where library functionality is re-implemented.

\Cref{tab:match} lists the applications we evaluated in this case study; all
were written in C or C++. Our methodology was as follows: first, function
implementations from the libraries in \Cref{tab:problems} were synthesized using
\presyn{}. Then, an existing tool \cite{Collie2019} was used to convert each of
the synthesized functions into a set of SMT constraints, which were then passed
to an off-the-shelf solver \cite{Ginsbach2018} to discover satisfying instances
in each of the applications. The number discovered in each application is given
in \Cref{tab:match}.
 Recent work \cite{Collie2020a} aims to use probabilistic synthesis for a similar
software engineering task but  does not evaluate
the performance of other synthesizers.

%Other work \cite{Collie2020a} aims to use probabilistic synthesis for a similar
%software engineering task. While this work provides a detailed evaluation of how
%synthesized code can be used to drive API migrations, it does not fully evaluate
%the performance of other synthesizers on their dataset.

\section{Related Work} \label{sec:related}
\paragraph{Sketching} One of the most important developments in program synthesis
is the idea of sketching \cite{Solar-Lezama2009, solar2013program}; it has been
used for a wide variety of purposes including auto-parallelization
\cite{Fedyukovich:2017:GSS:3062341.3062382} and SQL query generation
\cite{Wang2017}. Recent technical developments include  recursive tree
transformations \cite{Inala2017} and improved modularity of sketches
\cite{Singh2014}. Generally, such schemes require external sketch suggestions:
\presyn{} automates this by constructing priors over program corpora.
  
\paragraph{Types} Annotated types signatures or hints are often used to direct
program synthesis, most commonly for functional programs
\cite{Osera2015,Osera2019}. \textsc{Myth} \cite{Osera2015a} uses type signatures
alongside examples to synthesize recursive functional programs, while
\cite{Polikarpova2016} uses refinement types to guide the search process
\cite{Polikarpova2016}. In \cite{Collie2019a} extended type information is
suggested as a means of improving program synthesis, and in \cite{Collie2019} a
similar approach is used as a means of accessing heterogeneous accelerators for
scientific applications. Our work considers a wider, more diverse class of
libraries and applications and does not require human annotations or hints.

\paragraph{Neural Program Synthesis} The machine learning community has long
studied programming by example and input-output based program synthesis. Recent
work has examined both induction (with a learned latent version of the program)
and generation, which uses a language model to generate programs
\cite{Parisotto2016,allamanis2018survey, cummins2017synthesizing} 
%Such learned language models have been
%used for compiler-based benchmark generation \cite{cummins2017synthesizing},
%prediction of optimizations \cite{Cummins2017a} and compiler fuzzing
%\cite{Cummins2018}.

A recent trend in program synthesis is the use of neural networks and machine
learning. Generative approaches are focused on developing neural architectures
that interpret programs intrinsically \cite{Reed2015}. These approaches so far
struggle to generalize to large problem sizes and to complex semantic models
such as the compiler IR used by \presyn{}.

Others have used neural components to improve the performance of an existing
synthesizer. For example, both \cite{Balog2016} and \cite{Zohar2018a} aim to
learn from input-output examples; both require fixed-size inputs and outputs and
use a small DSL to generate training examples. Learned programs are limited to
list processing tasks; the DSLs targeted by these (and similar implementations
such as \textsc{SketchAdapt} \cite{Nye2019}) also rely on high level primitive
including (for example) primitives to tokenize strings or perform list
manipulations. 

\paragraph{White-Box} Our approach to synthesis (from \emph{black box} oracles)
is less widely studied than the corresponding \emph{white box} problem, where
the internal structure or implementation of a reference oracle is known.
Existing programs in specific problem domains are often used to synthesize
optimized implementations in a domain-specific language: the Halide
image-processing language \cite{Ragan-Kelley2013} is targeted by several
oracle-based approaches, based on \textsc{x86} assembly \cite{Mendis2015},
Fortran \cite{Kamil2016} and C++ \cite{Ahmad2019a} respectively. In
\cite{Cambronero2019}, abstract descriptions of software functionality are
learned based on dynamic traces and source code in order to facilitate
refactoring and other analyses.

Operating under the assumption of a black-box oracle means that many existing
approaches in program synthesis do not apply or fail to generalize to our
context \cite{Chen2019,Chen2019a}. By using a black-box oracle we are able to
avoid issues of generalization across datasets \cite{Parisotto2016a,Kalyan2018}.

%\paragraph{Synthesizing Imperative Programs} Prior work in imperative synthesis
%frequently focuses on straight-line code \cite{Gulwani2011a, Sasnauskas2017} or
%has to make special provision for control-flow \cite{Gulwani2011}. \simpl
%\cite{So2017} overcomes this problem by assuming a partial program is already
%provided (such as a loop structure).

\section{Conclusion} \label{sec:conclusion}
In this paper we have addressed the novel problem of \emph{black box} program
synthesis, where problem specifications are based on the observed behavior of an
existing component (rather than a human description). Our synthesizer,
\presyn{}, achieves better performance across a wide range of synthesis
benchmarks (composed of both new and existing problems) than five other
competing synthesizers.

As well as strong synthesis performance, the simple probabilistic models used to
implement \presyn{} provide interesting insights into the structure and
difficulty of synthesis problems. 
%Additionally, we demonstrate two case studies
%where \presyn{} can be used to solve software engineering problems.
 Our results
on two case studies are promising and we aim to pursue this further.
%this direction in the future.

%
%\begin{acks}
%  This work was supported by the \grantsponsor{epsrc}{Engineering and Physical
%  Sciences Research Council}{http://https://epsrc.ukri.org/} (grant
%  \grantnum{ppar}{EP/L01503X/1}), EPSRC Centre for Doctoral Training in
%  Pervasive Parallelism at the University of Edinburgh, School of Informatics.
%\end{acks}

%% Bibliography
% \bibliography{references,icsebibliography,pactreferences}
\bibliography{references}{}

%\appendix
%\section{Comparison of Synthesizers}
%\input{synths.tex}

\end{document}